\documentclass[final,3p,times,twocolumn]{elsarticle}
\usepackage{graphicx}
\usepackage{amsmath}
\usepackage{amssymb}
\usepackage{hyperref}
\usepackage{color}

\begin{document}

\title{Spin-phonon coupling in hole-doped pyrochlore iridates Y$_2$(Ir$_{1-x}$Ru$_x$)$_2$O$_7$: A Raman scattering study}

\author{Harish Kumar}
\address{School of Physical Sciences, Jawaharlal Nehru University, New Delhi - 110067, India.}
\author{V. G. Sathe}
\address{UGC-DAE Consortium for Scientific Research, University Campus, Khandwa Road, Indore - 452001, India.}
\author{A. K. Pramanik\corref{cor1}}
\ead{akpramanik@mail.jnu.ac.in}
\address{School of Physical Sciences, Jawaharlal Nehru University, New Delhi - 110067, India.}

\begin{abstract}
Temperature dependent Raman scattering measurements have been performed to explore unusual coupling between magnetism and crystal structure in doped pyrochlore iridate Y$_2$(Ir$_{1-x}$Ru$_x$)$_2$O$_7$ with $x$ = 0.0, 0.05 and 0.2. The parent Y$_2$Ir$_2$O$_7$ shows an onset of magnetic ordering around $\sim$ 160 K ($T_{N}$) which monotonically decreases with Ru doping. Further, magnetic moment also decreases with progressive substitution of Ru. Substitution of Ru$^{4+}$ (4$d^4$) for Ir$^{4+}$ (5$d^5$) does not introduce significant modification in structural parameters, however, the magnetic transition temperature decreases systematically with doping. Raman scattering data show an anomalous change in $A_{1g}$ and $P_3$ Raman mode frequency and line-width across $T_{N}$ of individual samples. We further show that the shifting of Raman mode frequency with temperature exhibits a strong deviation from anharmonic decay around and below the $T_{N}$ of respective samples which underlines a spin-phonon coupling in these materials.
\end{abstract}


\maketitle
\section {Introduction}
Iridium based pyrochlore materials (A$_2$Ir$_2$O$_7$, A = rare-earth elements) are of particular interest as they have been predicted to give rise many exotic topological phases.\cite{pesin,william} The 5$d$ based iridates are unique due to its heavy character and extended $d$-orbitals. The spin-orbital coupling (SOC) and electronic correlation ($U$) effect show a comparable energy scale. The interplay between SOC and $U$ is considered to give rise many interesting and novel physical properties. Moreover, pyrochlore materials have in-built frustration which arises from the structural geometry where the A and Ir cations form inter-penetrating corner-shared tetrahedra. This frustration, in general, again induces many other phenomena such as spin glass, spin ice, spin liquid, etc.\cite{gingras,yoshii,gardner1,nakatsuji,bramwell,fukazawa1} However, calculation has shown that inclusion of Dzyaloshinskii-Moriya interaction (DMI), which is very evident in SOC dominant systems, removes the spin degeneracy and favors the long-range ordered (LRO) state at low temperature in otherwise frustrated systems.\cite{elhajal}

The physical properties of pyrochlore iridates have been shown to evolve with A-site element: from magnetic and insulating to complex metallic and nonmagnetic phases with increasing ionic radii of A-ion.\cite{Matsuhira} This indicates that crystal structure particularly trigonal distortion of IrO$_6$ octahedra has crucial role in influence the physical properties which is directly related to the $x$-coordinate of O-atom ($X_p$-parameter). A recent theoretical study has predicted that this distortion is mainly associated with to F$_{2g}$ phonon modes.\cite{yang} Among the pyrochlore iridates, Y$_2$Ir$_2$O$_7$ is an ideal system because Y$^{3+}$ is nonmagnetic. Therefore, the physical properties are largely governed by Ir sublattice. The Y$_2$Ir$_2$O$_7$ shows insulating behavior throughout temperature and magnetic ordering around 160 K.\cite{shapiro,disseler,taira,fukazawa,soda,zhu,harish,kumar} The low temperature magnetic state of this material is, however, a debated issue. While the neutron measurements \cite{shapiro} show no sign of magnetic long-range ordering, previous muon spin relaxation investigations\cite{disseler} have shown spontaneous oscillations in muon asymmetry below 150 K, indicating a LRO state at low temperature. Theoretically, this material is proposed to be a Weyl-type semimetal (WSM) with a noncoplanar AFM structure.\cite{wan} Thus, Y$_2$Ir$_2$O$_7$ offers an ideal system to study the interplay between SOC and $U$ with the Ir-site substitution and to study the effect of structure distortion (trigonal) on the related properties.

With an aim to tune the SOC and $U$ parameters as well as to understand the role of structural distortion in Y$_2$Ir$_2$O$_7$, we have substituted Ru at Ir-site in Y$_2$(Ir$_{1-x}$Ru$_x$)$_2$O$_7$. Ru$^{4+}$ (4$d^4$) substitution for Ir$^{4+}$ (4$d^5$) in present series which act for hole doping and simultaneously it will tune both SOC and $U$, however, in opposite manner. Ru$^{4+}$ (4$d^4$) has comparable ionic radii with Ir$^{4+}$ (4$d^5$), thus no major structural modifications is expected in this series. Ru substitution in Y$_2$Ir$_2$O$_7$, previously, have shown the suppression of magnetic ordering and resistivity.\cite{kumar} Recently, Raman study has shown an interesting correlation between structure and magnetic behavior in Ir-based other compounds, such as, Sr$_2$IrO$_4$ and Na$_2$IrO$_3$.\cite{cetin,glama,gupta} There is exists Raman scattering studies on pyrochlore iridates,\cite{han,hasegawa} however, a direct correlation between magnetic and structure parameters is not established yet. Given that pyrochlore structure has complex structural organization and the Ir-based materials have reasonable SOC, which is weakened with Ru in present series, this present effort to investigate the magneto-structural coupling with Raman scattering measurements and the role of SOC would provide vital information in this class of materials.

In this paper, we present the results of magnetic and Raman scattering measurements for Y$_2$(Ir$_{1-x}$Ru$_{x}$)$_2$O$_7$ series with composition $x$ = 0.0, 0.05 and 0.2. Magnetic susceptibility data show that both the magnetic moment at low temperature and the magnetic irreversibility temperature ($T_{N}$) decrease with Ru substitution. With Ru doping, the overall system retains the original structural symmetry, however, the lattice parameters show an evolution with Ru content. Ru substitution reduces the trigonal distortion of IrO$_6$ octahedra in this series. Temperature dependent Raman scattering measurements for Y$_2$(Ir$_{1-x}$Ru$_{x}$)$_2$O$_7$ series show anomalous change in frequency and line width of Raman $A_{1g}$ and $P_3$ mode around magnetic ordering temperature ($T_{N}$) which suggest the presence of spin-phonon coupling. 

\begin{figure}[t!]
	\centering
		\includegraphics[width=7cm]{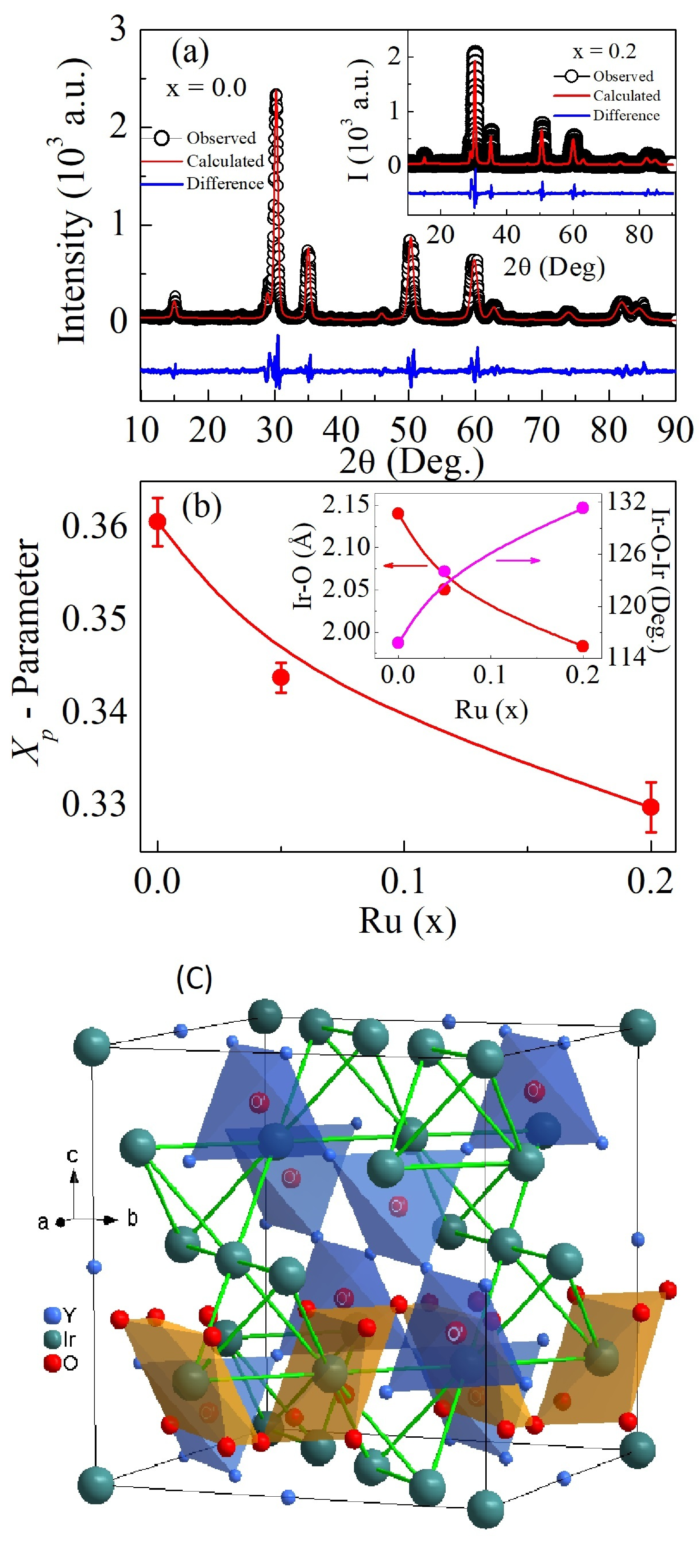}
	\caption{(a) XRD pattern of Y$_2$(Ir$_{1-x}$Ru$_x$)$_2$O$_7$ ($x$ = 0.0) collected at room temperature along with Rietveld refinement. Inset shows XRD pattern along with Rietveld refinement for $x$ = 0.2 composition. (b) $X_p$-parameter of O-atom is shown as a function of Ru doping for Y$_2$(Ir$_{1-x}$Ru$_x$)$_2$O$_7$ series. Inset shows Ir-O bond length and Ir-O-Ir bond angle as a function of Ru ($x$). (c) Cubic pyrochlore structure showing the Y$_4$O$'$ tetrahedra (Blue), Ir$_4$$\diamondsuit$ tetrahedra (Green) and IrO$_6$ octrahedra (light orange)}
	\label{fig:Fig1}
\end{figure}

\section {Experimental Details}
Series of polycrystalline materials Y$_2$(Ir$_{1-x}$Ru$_{x}$)$_2$O$_7$ ($x$ = 0.0, 0.05, and 0.2) have been prepared using conventional solid state method. Details of sample preparation and characterization are given in Ref.\cite{harish,kumar}. The phase purity of the samples have been checked using powder x-ray diffraction (XRD) using a Rigaku made diffractometer with Cu$K_\alpha$ radiation. The XRD data have been collected in the 2$\theta$ range of 10 - 90$^o$ at a step of $\Delta 2\theta$ = 0.02$^o$. The data have been refined using Reitveld refinement program (FULLPROF) for structural analysis by Young \textit{et al}.\cite{young} DC Magnetization ($M$) measurements have been carried out with a vibrating sample magnetometer (PPMS, Quantum Design). Temperature dependent Raman spectra have been collected using Diode based laser ($\lambda$ = 473 nm) coupled with a Labram-HR800 micro-Raman spectrometer. It is a single spectrometer with 1800 groves/mm grating and a peltier cooled CCD detector with a overall specral resolution of $\sim$ 1cm$^{-1}$. For the low temperature Raman measurements, the material has been mounted on a THMS600 stage from Linkam UK, with temperature stability of $\pm$ 0.1 K.

\section{Results and Discussions}
\subsection {Structural analysis}
Fig. 1 (a) shows the room temperature pattern of XRD data along with Rietveld analysis for the parent compound Y$_2$Ir$_2$O$_7$ ($x$ = 0.0). The XRD data for parent material match well with previous studies.\cite{shapiro,disseler,zhu,harish} The Rietveld refinement show the sample is in single phase and crystallizes in cubic \textit{Fd$\bar{3}$m} phase (Fig. 1 (a)) with lattice constant $a$ = 10.244(1) \AA. The inset of  Fig. 1 (a) shows the Rietveld refinement of XRD data for last sample of the present series i.e., $x$ = 0.2. It is clear in inset of Fig. 1 (a) that XRD pattern of $x$ = 0.2 does not exhibit any change with Ru content in terms of peak position or impurity peak which implies Ru doping does not cause any major changes in crystal symmetry. This behavior is also expected for Ru doping because of their matching ionic radii (Ru$^{4+}$ = 0.62 \AA and Ir$^{4+}$ = 0.625 \AA). However, the Rietveld analysis of XRD data for all samples shows that structural parameters ($X_p$-parameter, bond-length and bond-angle) evolve continuously with Ru doping in the present series as shown in Fig. 1 (b). Fig. 1 (c) depicts unit cell atomic arrangement of cubic-\textit{Fd$\bar{3}$m} pyrochlore structure for the parent compound Y$_2$Ir$_2$O$_7$ showing it has interpenetrating two types of sub-lattices of tetrahedra, i.e., Y$_4$O$'$ and Ir$_4$$\diamondsuit$, where $\diamondsuit$ implies empty center-site (8a site). As Y$^{3+}$ is nonmagnetic in parent case so the tetradehra of Ir cations only introduces magnetic interaction/frustration in Y$_2$Ir$_2$O$_7$.

Pyrochlore materials, generally, adopt the cubic crystallographic phase with \textit{Fd$\bar{3}$m} space group.\cite{subram} There are two free structural parameters in pyrochlore structure such as, lattice constant ($a$) and $x$-coordinate of oxygen 48$f$-site ($X_p$-parameter). Apart from the lattice parameters, $X_p$-position of the oxygen is directly related to the structural organization of IrO$_6$ octahedra in the pyrochlore structure. In general, the six oxygen ions are coordinated around Ir$^{4+}$ cation to form a distorted IrO$_6$ octahedra. The amount of (trigonal) distortion of the IrO$_6$ octahedra can be determined from the information of the $X_p$-parameter of oxygen atom. In IrO$_6$ octahedra, each oxygen ion is at equal distance from central Ir ion. For $x_{ideal}$ = 0.3125, Ir$^{4+}$ cations form an ideal IrO$_6$ octahedra and if $x$ = 0.375, the IrO$_6$ octahedra are distorted where A cations form a perfect cube. The pyrochlore structure consists of two types inter-penetrating layer of tetrahedra i.e., Y$_4$O$'$ and Ir$_4$$\diamondsuit$, where $\diamondsuit$ implies empty center-site (8a site) as illustrated in Fig. 1 (c). All pyrochlore iridates usually have larger $X_p$-value than this ideal value, indicating trigonal distortion to the octahedra which lowers the cubic symmetry and gives trigonal crystal field.\cite{yang,JP,Hozo}

\begin{figure}
  \centering
		\includegraphics[width=7cm]{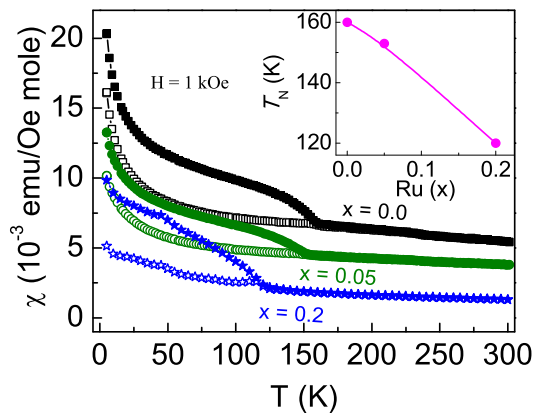}
  \caption{Temperature dependent magnetic susceptibility data measured in 1 kOe field under ZFC and FC protocol are shown for Y$_2$(Ir$_{1-x}$Ru$_{x}$)$_2$O$_7$ series. Inset shows variation of transition temperature $T_{N}$ with Ru substitution for the same series.}
  \label{Fig.2}
\end{figure}

The variation in $X_p$-parameter of oxygen atom with Ru is shown in Fig. 1 (b). The $X_p$-value for the parent system is found around 0.36 which is large compared to the ideal value ($x_{ideal}$ = 0.3125) and suggests the IrO$_6$ octahedra's are distorted and compressed. As evident in figure, the $X_p$-value has decreased with progressive substitution of Ru content. The decrease of the $X_p$-value in Fig. 1 (b) suggests that the system is leading towards perfect octahedra, hence trigonal crystal field is reduced. As the oxygen atoms occupying the 48$f$-site form the IrO$_6$ octahedra, hence, the $<$Ir-O-Ir$>$ bond-angle and Ir-O bond-length are related to the $X_p$-parameter and plays an important role in determining the physical properties. The evolution of Ir-O bond-length and $<$Ir-O-Ir$>$ bond-angle for the present series are shown inset of Fig. 1 (b). It is evident in figure that the bond-length decreases while bond-angle simultaneously increases with progressive substitution of Ru. This variation of Ir-O bond-length and angle suggests that distortion in IrO$_6$ octahedra is reduced and increases Ir(5\textit{d})/Ru(4\textit{d}) and O(2\textit{p}) orbital overlapping which facilitates the hopping of the Ir electrons.

\subsection{Magnetization Study}
Fig. 2 shows the temperature dependent magnetic susceptibility ($\chi$) data for Y$_2$(Ir$_{1-x}$Ru$_x$)$_2$O$_7$ series, where the data have been collected under zero-field cooled (ZFC) and field-cooled (FC) protocol in an applied field of 1 kOe. In Fig. 2, it is quite obvious that $\chi(T)$ data for parent material ($x$ = 0.0) shows a magnetic irreversibility around $T_{N}$ = 160 K which marks an antiferromagnetic type magnetic phase transition where below this temperature a large bifurcation between the $\chi_{ZFC}$ and $\chi_{FC}$ branches is observed. On further cooling, we find that below around 50 K the magnetization increases very sharply. As evident in Fig. 2, with Ru substitution the notable observations are, \romannumeral 1) the low temperature magnetic moment decreases, \romannumeral 2) difference between $\chi_{FC}$ and $\chi_{ZFC}$ decreases and \romannumeral 3) the bifurcation temperature ($T_{N}$) shift toward low temperature. The composition dependent $T_{N}$ is shown in inset of Fig. 2. Moreover, the decrease of $T_{N}$ is almost linear with Ru substitution and we find that magnetic irreversibility temperature $T_{N}$ $\sim$ 120 K for 20\% doping of Ru ($x$ = 0.2) which is quite consistent with previous study.\cite{kumar}

It can be further observed in Fig. 2 that the magnetic susceptibility decreases with increasing $x$ which is rather surprising as data suggest substitution of Ru$^{4+}$ (S = 1) for Ir$^{4+}$ ($J_{eff}$ = 1/2) causes weakening and suppression of magnetic transition in Y$_2$(Ir$_{1-x}$Ru$_x$)$_2$O$_7$. Generally, the pyrochlore lattice with Heisenberg-type nearest neighbor interactions show collinear type antiferromagnetic (AFM) spin structure where frustration is a very inherent feature. An additional Dzyaloshinskii-Moriya (DM) interaction, which mainly originates from the spin-orbit coupling (SOC), results in non-collinear type magnetic structures.\cite{elhajal} In present case, the introduction of Ru in place of Ir is expected to reduce the SOC effect, hence it would derive toward the system with more collinear type AFM spin ordering. This can lead to decrease of moment. Moreover, recent theoretical calculations have shown that with increasing $U$ the magnetic state in Y$_2$Ir$_2$O$_7$ moves from non-collinear AIAO state to AFM spin structure.\cite{shinaoka,Ishii} Therefore, the decrease in magnetic moment may be due to the tuning of SOC and $U$ with Ru doping. In Fig. 2, therefore, magnetic susceptibility data in comparison to $x$ = 0.1 the data for $x$ = 0.2 look significantly reduced. Here, it can be further mentioned that the other end member of this present series, i.e., Y$_2$Ru$_2$O$_7$ ($x$ =1.0) shows an antiferromagnetic transition ($T_N$ $\sim$ 76 K) and large bifurcation between ZFC and FC magnetization at low temperature.\cite{Kmiec} Additionally, this materials also shows reasonably large frustration which suggests Y based pyrochlore materials have inherent frustration. 

\begin{figure}[t!]
  \centering
		\includegraphics[width=7cm]{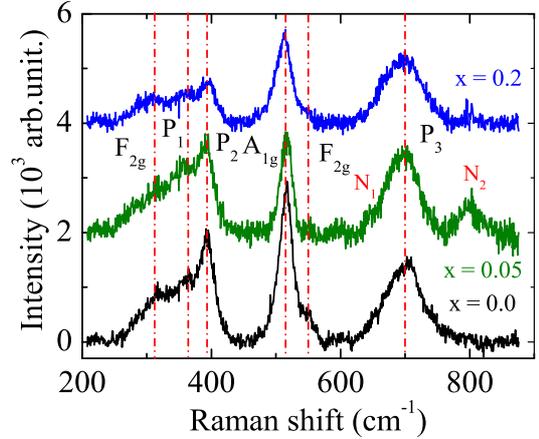}
  \caption{Raman spectra measured at room temperature are shown for Y$_2$(Ir$_{1-x}$Ru$_x$)$_2$O$_7$ ($x$ = 0.0, 0.05 and 0.2). Raman data for $x$ = 0.05 and 0.2 are shifted vertically for clarity.}
  \label{Fig.3}
\end{figure}

\begin{figure}[t!]
  \centering
		\includegraphics[width=7cm]{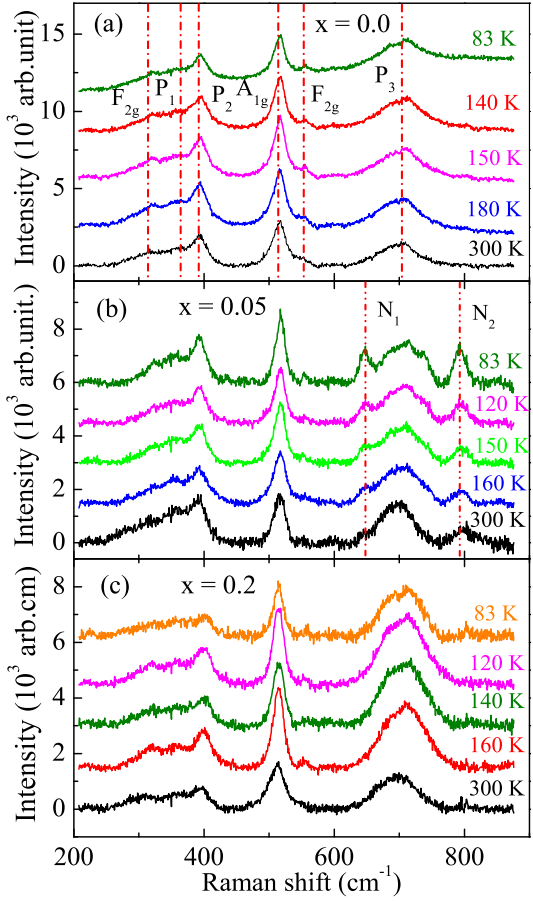}
  \caption{Raman spectra have been shown for $x$ = 0.0 (a), $x$ = 0.05 (b) and $x$ = 0.2 (c) for Y$_2$(Ir$_{1-x}$Ru$_x$)$_2$O$_7$ series at selected temperatures.}
  \label{Fig.4}
\end{figure}

\subsection{Temperature dependence Raman study for Y$_2$Ir$_2$O$_7$}
To understand further the structural evolution, Raman spectroscopy has been carried out which are primarily sensitive to metal ion-oxygen bond vibrations as well as to local structural disorder. Fig. 3 shows Raman spectra for the Y$_2$(Ir$_{1-x}$Ru$_x$)$_2$O$_7$ series with composition $x$ = 0.0, 0.05 and 0.2 at room temperature. According to previous structural discussions, Y$_2$Ir$_2$O$_7$ belongs to the cubic space group of \textit{Fd$\bar{3}$m}. For this space group, six Raman active modes are expected according to the factor group analysis, $\Gamma^R$ = $A_{1g}$ + E$_g$ + 4F$_{2g}$ which are mainly involved in motion of oxygen atoms.\cite{han,hasegawa} At room temperature, we have observed six Raman peaks at 314, 361, 393, 516, 552, and 700 cm$^{-1}$ which are assigned to $F_{2g}$, $P_1$, $P_2$, $A_{1g}$, $F_{2g}$ and $P_3$ Raman modes, respectively. The observed modes for Y$_2$Ir$_2$O$_7$ are fully consistent with the previous reported studies of the pyrochlore iridates.\cite{han,hasegawa} Out of all, the $A_{1g}$ mode (515 cm$^{-1}$) has particular importance in the pyrochlore structure because it is directly involved in the trigonal distortion of IrO$_6$ octahedra through modification of $<$Ir-O-Ir$>$ bond angle.\cite{Taniguchi} Therefore, $A_{1g}$ mode is involved in the modulation of oxygen position coordinate ($X_p$) where the vibration of oxygen atoms at 48$f$ site plays a major role.\cite{Sanjuan,tracy} Similarly, F$_{2g}$ mode is assigned to the mixture of A-O and Ir-O bond stretching vibrations and E$_g$ mode corresponds to IrO$_6$ bending vibrations.\cite{sreena} Therefore, both $A_{1g}$ and $E_g$ modes are basically considered as bending mode and largely influenced by changes with $<$Ir-O-Ir$>$ bond angle.

The $P_3$ phonon mode in pyrochlore materials, on the other hand, is another vital parameter which is mostly observed around 700 cm$^{-1}$ and is related to Ir-O bond stretching which has significant influence on the Ir(5$d$)-O(2$p$) hybridization.\cite{Granado1,Mihal} With Ru substitution, all the six modes are observed in doped samples. In addition, $x$ = 0.05 material shows two extra peaks at 648, and 792 cm$^{-1}$ which are marked as N$_1$ and N$_2$ modes, respectively. The N$_2$ mode is also observed for $x$ = 0.2 but intensity of N$_2$ mode for $x$ = 0.2 is very low. The observed new modes are commonly found in the frequency range from 600 to 900 cm$^{-1}$ in pyrochlores \cite{Sanjuan,tracy} and are often assigned as the second order Raman scattering.\cite{hasegawa,Saha}  
In order to understand an evolution of structural symmetry with temperature in the parent and doped materials, we have performed the temperature dependent Raman measurements. Fig. 4 (a) shows the representative Raman spectrum of Y$_2$Ir$_2$O$_7$ at 300, 180, 150, 140 and 83 K where the temperatures correspond to the room temperature paramagnetic state (300 K), just above (180 K) and below (140 K) the magnetic transition temperature $T_{N}$ (160) and low-temperature state (83 K). It is evident in the figure that there is no noticeable changes in Raman active modes in terms of shifting of peak position or appearance of new peak with lowering the temperature. This implies that there is no visible change in structural symmetry with temperature which agrees well with the temperature dependence of XRD and neutron scattering measurements.\cite{harish, shapiro} 

\begin{figure}[t!]
  \centering
		\includegraphics[width=8cm]{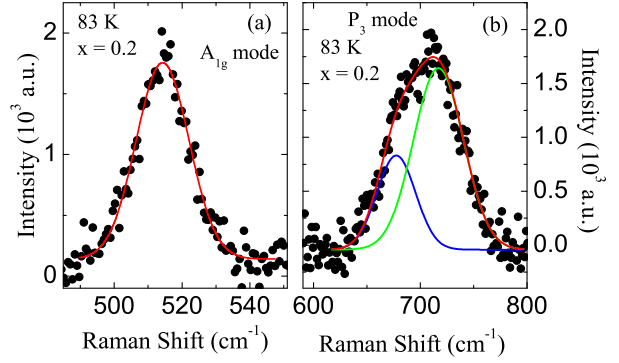}
  \caption{(a) and (b) show lineshape of $A_{1g}$ and $P_{3}$ phonon modes at temperature 83 K for $x$ = 0.2 sample, respectively. The solid red line in figure for $A_{1g}$ and $P_{3}$ modes is due to single and multiple peak fitting.}
  \label{Fig.5}
\end{figure}
 
\begin{figure}[t!]
  \centering
		\includegraphics[width=8cm]{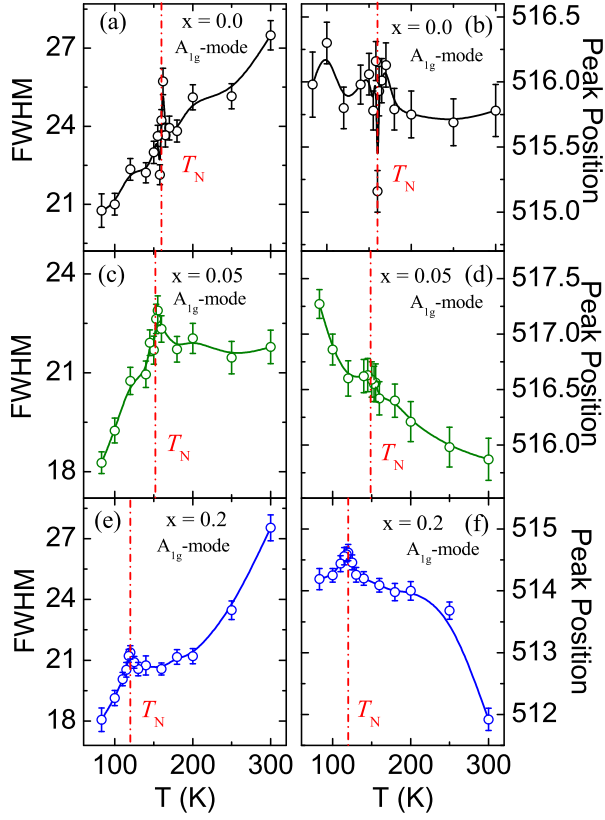}
  \caption{(a), (c), (e) show full width at half maximum (FWHM), and (b), (d), (f) show position/frequency of the $A_{1g}$ phonon mode as a function of temperature for $x$ = 0.0, 0.05 and 0.2 samples at upper, middle and lower panel, respectively.}
  \label{Fig.6}
\end{figure}

As seen in Fig. 1 (b), all structural lattice parameters show an evolution with doping content ($x$). Therefore, we now turn to the temperature dependent Raman study for the doped samples. Figs. 4 (b) and (c) show temperature dependent Raman spectrum for the doped $x$ = 0.05 and 0.2 samples, respectively. With lowering the temperature, no changes in low frequency modes are observed for 5\% doped sample but the $A_{1g}$ (515 cm$^{-1}$) and the new N$_1$ and N$_2$ modes become more sharp and intense. For 20\% doped compound, the intensity of $A_{1g}$ mode increases with decreasing temperature. To understand further the role of phonon parameters with temperature, we have performed detail analysis of temperature dependent Raman data for all the samples.

The $A_{1g}$ (515 cm$^{-1}$) and $P_{3}$ (700 cm$^{-1}$) mode play an important role in pyrochlore structure. We have performed detail analysis of these two modes in order to understand the their evolution with doping as well as with temperature, particularly around the magnetic transition temperature ($T_{N}$). As evident in Fig. 4 that the $A_{1g}$ peaks are single and sharp while the $P_{3}$ peaks are broad and consist of multiple peaks. Therefore, we have performed single and multiple peak fitting for $A_{1g}$ and $P_{3}$ modes, respectively. The representative peak fitting for these $A_{1g}$ and $P_{3}$ modes is shown in Figs. 5a and 5b, respectively for $x$ = 0.2sample at 83 K. The evolution of line-width as represented by full width at half maximum (FWHM) and position of $A_{1g}$ phonon mode with temperature are shown for the parent material $x$ = 0.0 as shown in Fig. 6 (a) and (b), respectively. With lowering the temperature, frequency/position of $A_{1g}$ phonon mode show fluctuations. This mode is directly associated with the oxygen $X_p$-parameter of IrO$_6$ octahedra therefore, the observed fluctuations with temperature in the frequency of $A_{1g}$ mode are in good agreement with the previously reported temperature dependent of $X_p$-parameter obtained from XRD analysis.\cite{harish} The change in line-width is about -24.5\% over the temperature range. It is important to note that a discernible anomaly in FWHM of $A_{1g}$ phonon mode is observed around $T_{N}$. This anomaly in Raman data may arise due to sudden rotation/distortion of IrO$_6$ octahedra across $T_{N}$ temperature (160 K). 

\begin{figure}[t!]
  \centering
		\includegraphics[width=8cm]{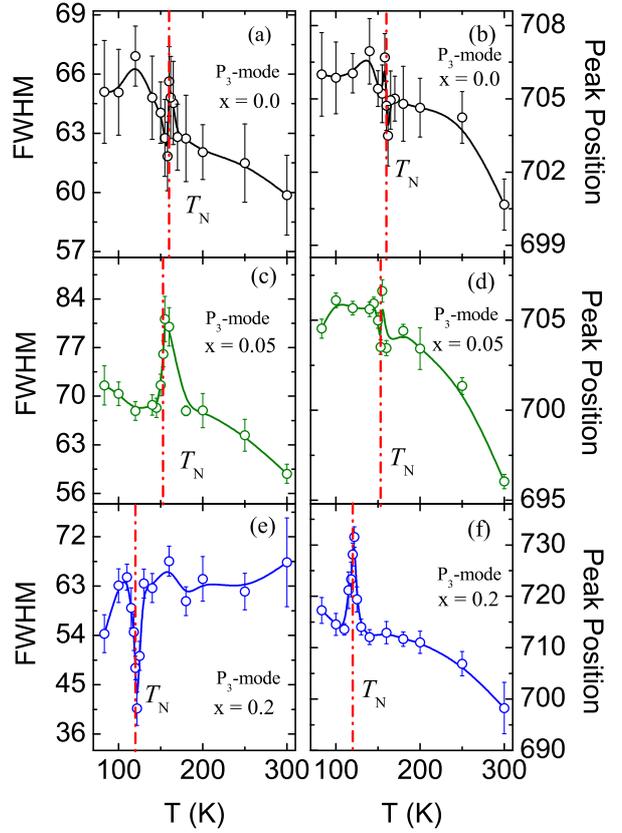}
  \caption{(a), (c), (e) show full width at half maximum (FWHM), and (b), (d), (f) show position/frequency of the $P_{3}$ phonon mode as a function of temperature for $x$ = 0.0, 0.05 and 0.2 samples at upper, middle and lower panel, respectively.}
  \label{Fig.7}
\end{figure}

For Ru doped materials, Figs. 6 (c) and (d) depict the FWHM and the frequency of the 515 cm$^{-1}$ $A_{1g}$-mode as function of temperature for $x$ = 0.05 sample. Fig. 6 (c) shows the FWHM of the $A_{1g}$ mode decreases with the lowering temperature where the change in FWHM is $\sim$ -16.2\% over this temperature regime. It is important to note that FWHM of $A_{1g}$ mode show a clear anomaly around 152 K which is corresponds to the magnetic ordering temperature $T_{N}$ of $x$ = 0.05 material. The change in $A_{1g}$ mode position with temperature is presented in Fig. 6 (d) for $x$ = 0.05 which increases with decreasing the temperature and shows a distinct anomaly around $T_{N}$ ($\sim$ 152 K). This change in position of $A_{1g}$ mode is not substantial, as we calculate the change is about 0.16\% over this temperature region. Compared to parent material ($x$ = 0.0), the peaks in both FWHM and position of $A_{1g}$ mode is shifted by 8 K for $x$ = 0.05 which is consistent with a suppression of the magnetic ordering as presented in Fig. 2. 

With further increasing the doping concentration to $x$ = 0.2, Fig. 6 (e) shows the FWHM of the $A_{1g}$ mode which decreases with decreasing temperature where the change in FWHM is $\sim$ -34.4\% over this temperature regime. Moreover, for the $x$ = 0.2, Fig. 6 (f) shows the temperature evolution of the position of $A_{1g}$ mode which increases with decreasing the temperature. We calculate the change in position of $A_{1g}$ is about 0.4\% ($\sim$ 2 cm$^{-1}$) over whole temperature region. We further observe a change in phonon frequency of $A_{1g}$ modes around 4 cm$^{-1}$ for $x$ = 0.2 sample at room temperature compared to parent material. The position of $A_{1g}$ mode also shows a distinct anomalies around $\sim$ 120 K which is corresponding to magnetic ordering temperature ($T_{N}$) of $x$ = 0.2. The observed anomalies in the FWHM and the position of $A_{1g}$ mode around $T_{N}$ for both the doped material ($x$ = 0.05, 0.2) is in good agreement with suppression of the magnetic ordering with Ru content ($x$). The previous Raman study for the other iridate systems Sr$_2$(Ir$_{1-x}$Ru$_{x}$)O$_4$ reveals similar evolution of phonon parameters around the magnetic ordering temperature.\cite{cetin,glama} The observed anomaly in Raman data for the present series may arises due to sudden rotation/distortion of IrO$_6$ octahedra across $T_{N}$.

\begin{figure}
  \centering
		\includegraphics[width=8.5cm]{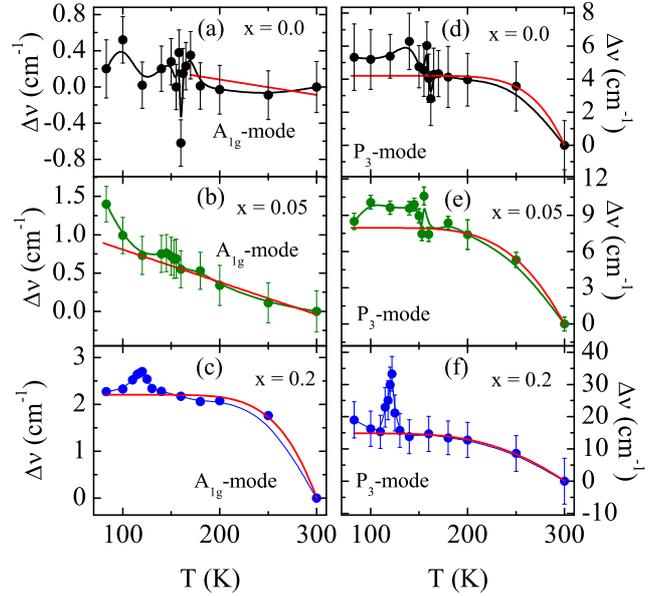}
  \caption{(a), (c), (e) show temperature evolution of the $A_{1g}$ phonon frequency shift and (b), (d), (f) for  the $P_{3}$ phonon frequency shift for $x$ = 0.0, 0.05 and 0.2 samples at upper, middle and lower panel, respectively. Solid lines are fit corresponding to anharmonicity predictions (Eq. 1).}
  \label{Fig.8}
\end{figure}

We have earlier discussed the importance of the $P_3$ phonon mode which is observed around 700 cm$^{-1}$. Here, we have analyzed the temperature dependent Raman data in order to understand the evolution of the $P_{3}$-phonon mode across magnetic ordering temperature in the present series. Fig 7 (a), (c), (e) and (b), (d), (f) show the evolution of FWHM and frequency/position of $P_{3}$-phonon mode with temperature for the present Y$_2$(Ir$_{1-x}$Ru$_x$)$_2$O$_7$ series, respectively. We find that the FWHM of $P_3$ phonon mode for $x$ = 0.0 and 0.05 increases with decreasing the temperature and shows a distinct anomaly around $T_{N}$. While, the change in FWHM of $P_3$ phonon mode for $x$ = 0.2 is small but it exhibits a downfall anomaly around $T_{N}$. The position of $P_3$ phonon mode for $x$ = 0.0, 0.05 and 0.2 mostly increases with decreasing the temperature in contrast to $A_{1g}$ mode. The observed phonon anomalies in terms of peak position and FWHM of $P_{3}$ mode around $T_{N}$ for all the materials is in good agreement with suppression of the magnetic ordering temperature $T_N$ with Ru content ($x$).

With material composition ($x$), the phonon frequency of $P_3$ mode also modifies. For $x$ = 0.2, we find the change in phonon frequency of $P_3$ about 3 and 22 cm$^{-1}$ at room temperature and magnetic transition temperature, respectively compared to $x$ = 0 material. Given that this change of phonon frequency represents a modification of electronic and local structures, these results imply that the Ru has nonuniform effect on shifting of phonon frequency for $A_{1g}$ and $P_3$ modes. Nonetheless, these changes are comparable with other iridate samples. For instance, in case of Sr$_2$Ir$_{1-x}$Ru$_x$O$_4$ the change is around 8 cm$^{-1}$ for $x$ = 0.2 sample.\cite{glama}

\begin{figure}
  \centering
		\includegraphics[width=6.5cm]{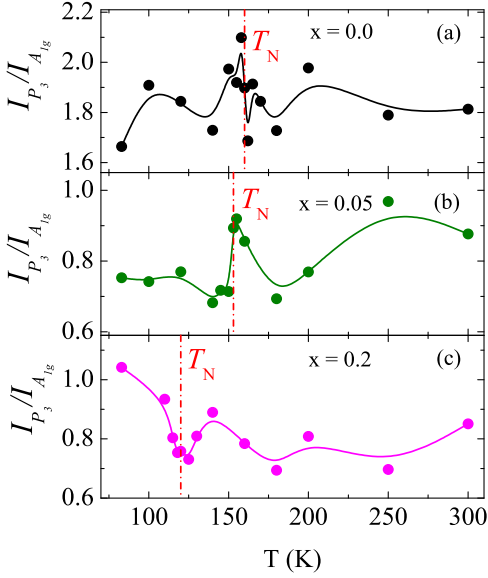}
  \caption{(a), (b) and (c) show temperature dependence of the relative intensity of the $A_{1g}$ to the $P_{3}$ phonon mode for $x$ = 0.0, 0.05 and 0.2 samples, respectively.}
  \label{Fig.9}
\end{figure}

Above discussions imply that the anomalous changes in position and in line-width (FWHM) of the Raman mode frequency around the magnetic ordering temperature mainly occur due to spin-phonon coupling in these systems. To understand further the presence of spin-phonon coupling in present series, we have analyzed the Raman data of present series with the following anharmonic decay model\cite{Vermette}

\begin{eqnarray}
\omega(T) = \omega_0 - C \left[1 + \frac{2}{\left[\exp\left(\frac{\hbar \omega_0}{2K_{B}T}\right) -1\right]}\right]
\end{eqnarray}
 
where $\omega_0$ and $C$ are the intrinsic frequency of the optical mode and anharmonic coefficient, respectively. The term $[\exp(\frac{\hbar \omega_0}{2K_{B}T}) -1]^{-1}$ corresponds to the thermal population factor of the acoustic modes. In general, temperature dependent change of phonon mode frequency is observed to follow anharmonic two-channel-phonon decay which is given in Eq. 1 (assuming negligible contribution from lattice expansion/contraction and/or electron-phonon coupling term). However, due to presence of prominent spin-phonon coupling which is caused by change in exchange interaction by lattice vibration, the phonon mode frequency shift shows a deviation from anharmonic decay around the magnetic phase transition. Granado \textit{et al}\cite{Granado} has proposed a model for spin-phonon coupling showing the contribution can be expressed as,

\begin{eqnarray}
  \omega_{s-ph}(T) = \lambda \left\langle S_i.S_j \right\rangle
\end{eqnarray}

where $\lambda$ is spin-phonon coupling constant and $\left\langle S_i.S_j \right\rangle$ is spin correlation function. At low temperatures and particularly near to the magnetic phase transition, the spin-spin correlation develops, hence its contribution become prominent showing a marked deviation from Eq. 1. As spin-spin exchange interaction is key to spin-phonon coupling, spin-orbit coupling which is prominent in Ir-based samples also plays a vital role in exchange coupling.

We have shown the shifting of phonon mode frequency $\Delta\nu$ (normalized to room temperature) for $A_{1g}$ and $P_3$ modes in Fig. 8 (a), (b), (c) and (d), (e), (f), respectively for Y$_2$(Ir$_{1-x}$Ru$_x$)$_2$O$_7$ series. The solid lines in Fig. 8 are due to fitting with Eq. 1. It is clear in figure that for parent compound ($x$ = 0.0), the frequency shifting ($\Delta\nu$) can be roughly fitted in high temperature region but below and around $T_{N}$ = 160 K, the $\Delta\nu$ shows fluctuations similar to the position of $A_{1g}$ mode (Fig. 6 (b)). We have discussed before that the $A_{1g}$ phonon mode corresponds to trigonal distortion of IrO$_6$ octrahedra through the vibration of O-atom. Therefore, this mode is directly related to the oxygen $X_p$-parameter of IrO$_6$ octahedra. The observed fluctuations with temperature in the $A_{1g}$ phonons are in good agreement with the previously reported temperature dependent $X_p$-parameter obtained from XRD analysis.\cite{harish} However, with increasing Ru concentration, the high temperature frequency shift ($\Delta\nu$) can be fitted well with Eq. 1. It is noticeable in figure that $\Delta\nu$ strongly deviates from fitting of Eq. 1 across the magnetic ordering temperature ($T_{N}$).

In case of $P_3$ mode, the $\Delta\nu$ shows a large deviation from Eq. 1 across $T_{N}$ in parent compound. This deviation from Eq. 1 in Fig. 8 (d) underlines the fact that this has prominent spin-phonon coupling. With Ru substitution, the $\Delta\nu$ deviation for $P_3$ mode shows a pronounced softening across $T_{N}$ of respective samples (Figs. 8e and 8f). It is interesting that deviation of $P_3$ mode reduces with Ru substitution while through Ru$^{+4}$ contributes to magnetic ordering. We speculate that the tuning of SOC with Ru doping modifies the spin-spin interaction which may lead to softening of $\Delta\nu$ deviation across $T_{N}$ in higher doped samples. It is well known fact that SOC induce distortion of IrO$_6$ octahedra induces Dzyaloshinskii-Moriya (DM) type asymmetric interaction which subsequently results in a weak ferromagnetic behavior in Ir-based Sr$_2$IrO$_4$.\cite{crawford,imtiaz} This material otherwise exhibits Heisenberg-type antiferromagnetic exchange interaction with magnetic transition temperature $\sim$ 240 K.\cite{imtiaz} In fact, we observe a huge change in Ir-O-Ir bond angle from around 115 to 131$^o$ with 20\% of Ru doping while both Ir$^{4+}$ and Ru$^{4+}$ have matching ionic radii (see Fig. 1(b)). Therefore, softening of deviation in frequency shifting of $P_3$ mode in Fig. 8 is likely to be associated with tuning of SOC in Y$_2$(Ir$_{1-x}$Ru$_x$)$_2$O$_7$ series.

We have further looked at the ratio of $A_{1g}$ and $P_3$ phonon modes in present series which are related to Ir-O bond angel and bond length, respectively (discussed above). We have plotted the ratio of $P_3$ and $A_{1g}$ phonon intensity mode as a function of temperature for $x$ = 0.0, 0.05 and 0.2 in Fig. 9. For parent material ($x$ = 0.0), with decreasing the temperature the $I_{P_3}$/$I_{A_{1g}}$ shows similar type of fluctuations as observed for position of $A_{1g}$ phonon mode (Fig. 6 (b) and Fig. 8 (a)). The intensity ratio of $I_{P_3}$/$I_{A_{1g}}$ for $x$ = 0.05 shows a small variation with temperature, however, it shows an anomalous change around the magnetic ordering temperature ($T_{N}$). For the highest doped material i.e., $x$ = 0.2, the intensity ratio of $I_{P_3}$/$I_{A_{1g}}$ shows a distinct anomaly around $T_{N}$. Further, we have separately analyzed the temperature dependent intensity data of $A_{1g}$ and $P_3$ mode for $x$ = 0.2 (not shown). While both $A_{1g}$ and $P_3$ modes show an anomalous change around $T_{N}$ but the change in $A_{1g}$ mode is comparatively high. 

Our Results show that Ru substitution in Y$_2$(Ir$_{1-x}$Ru$_x$)$_2$O$_7$ series has prominent effects on structural behavior where the system retains its original structural symmetry but the phonon modes show an evolution with temperature. The anomalous changes in frequency and line-width of the Raman modes around the magnetic ordering temperature are quite interesting. Further, the anomaly in $I_{P_3}$/$I_{A_{1g}}$ ratio and deviation in frequency shifting from anharmonic decay across $T_{N}$ for both $I_{P_3}$ and $I_{A_{1g}}$ modes (Fig. 9) suggest a reasonable spin-phonon coupling in present Y$_2$(Ir$_{1-x}$Ru$_x$)$_2$O$_7$. These results in present study will certainly be helpful to understand the low temperature nature of structural behavior with the magnetic behavior in pyrochlore iridates.
 
\section{Conclusion}
To summarize, we have presented Raman scattering study on series of polycrystalline samples Y$_2$(Ir$_{1-x}$Ru$_x$)$_2$O$_7$ ($x$ = 0.0, 0.05 and 0.2). With substitution of Ru, system retains its original crystal structure symmetry, however, structural parameters evolve with the Ru doping ($x$). Magnetic susceptibility data reveal both the onset of magnetic irreversibility and the magnetic moment decreases with Ru doping. Temperature dependent Raman data of parent material Y$_2$Ir$_2$O$_7$ does not exhibits any structural change around magnetic transition $T_{N}$, but Raman mode frequency and line-width show an anomalous change around the magnetic transition temperature $T_{N}$. With Ru substitution, the frequency and line-width of $A_{1g}$ and $P_3$ phonon mode also exhibit an anomaly around $T_{N}$ of respective samples. Shifting of Raman mode frequency with temperature shows a deviation from anharmonic decay model around respective magnetic ordering temperature which signifies prominent spin-phonon coupling in these samples. Tuning of SOC with Ru substitution is also evident in softening of frequency deviation from anharmonic decay.

\section{Acknowledgment} 
We acknowledge UGC-DAE CSR, Indore for the magnetization and Raman spectroscopy measurements. We are thankful to Dr. Alok Banerjee for the magnetization measurements and discussions. We are thankful to Mr. Kranti Kumar and Ajay Kumar Rathore for the helps in measurements. HK acknowledges UGC, India for BSR fellowship.

\end{document}